\documentclass[12pt]{article}
\usepackage{cite,psfig}

\def\be{\begin{equation}}
\def\ee{\end{equation}}
\def\bea{\begin{eqnarray}}
\def\eea{\end{eqnarray}}
\def\epsilon{\varepsilon}

\begin{document}
\thispagestyle{empty}
\vspace*{1cm}
\noindent
{\hspace*{\fill} CERN-TH/2001-373

\vspace*{2cm}

\begin{center}
{\Large\bf 5D Super Yang-Mills Theory in 4D Superspace,\\[.3cm]
Superfield Brane Operators,\\[.5cm] 
and Applications to Orbifold GUTs}
\\[2.5cm]
{\large A. Hebecker
}\\[1.0cm]
{\it Theory Division, CERN, CH-1211 Geneva 23, Switzerland}
\\[.2cm]
(December 17, 2001)
\\[2.5cm]

{\bf Abstract}\end{center}
\noindent
A manifestly gauge invariant formulation of 5-dimensional supersymmetric 
Yang-Mills theories in terms of 4d superfields is derived. It relies on a 
supersymmetry and gauge-covariant derivative operator in the $x^5$ direction. 
This formulation allows for a systematic study of higher-derivative 
operators by combining invariant 4d superfield expressions under the 
additional restriction of 5d Lorentz symmetry. In cases where the 5d 
theory is compactified on a gauge-symmetry-breaking orbifold, the formalism
can be used for a simple discussion of possible brane operators invariant 
under the restricted symmetry of the fixed points. This is particularly 
relevant to recently constructed grand unified theories in higher dimensions 
(orbifold GUTs). Several applications, including proton decay operators and 
brane-localized mass terms, are discussed. 
\newpage

\section{Introduction}
The standard framework for the discussion of physics above the electroweak
scale is supersymmetric grand unification. Taking the phenomenological 
success of traditional gauge coupling unification seriously, the energy 
range between the GUT scale and the string (or Planck) scale is the natural 
domain for higher-dimensional field theories. Indeed, starting with the 
proposal of Kawamura~\cite{kaw}, a number of very simple and realistic 
higher-dimensional GUT models have recently been constructed~\cite{af,hn,kk,
hm1,hm2,abc,hnos,hjll,cprt,hks,oth}. Of course, numerous other interesting 
ideas that are based on supersymmetry (SUSY) in higher dimensions exist, for 
example, the extra-dimensional SUSY breaking scenarios of~\cite{ant} or the 
intermediate scale unification models of~\cite{ddg}. 

Both conceptually and for the discussion of low-energy phenomenology, a 
4d superfield description of the higher-dimensional SUSY is desirable. 
After the early work of~\cite{mss}, this issue has recently been revived 
in~\cite{agw,mpo}. In the present paper, we develop the formalism
of~\cite{agw,mpo} by providing a manifestly gauge-invariant 4d superfield 
description of the non-abelian 5d theory. Our particularly simple 
formulation relies on the consistent use of the supersymmetry- and 
gauge-covariant derivative operator in the $x^5$ direction. Furthermore, we 
find that, starting from conventional 4d Super Yang-Mills (SYM) theory and 
introducing $x^5$ (together with the corresponding gauge connection) 
as an additional parameter, the full 5d supersymmetric theory is 
unambiguously determined. This approach, which is based on combining 
invariant 4d superfield expressions under the additional restriction of 5d 
Lorentz symmetry, can also be used for a systematic study of higher 
derivative terms in the 5d SYM theory.

The presented gauge covariant formulation allows for a simple discussion of 
superfield brane operators invariant under the restricted gauge symmetry 
of orbifold fixed points. This is particularly relevant to the 
extra-dimensional GUT models mentioned earlier, where the GUT symmetry is 
broken to the standard model (SM) gauge group by the boundary conditions 
of an orbifold compactification. More specifically, SU(5) models in 
5 dimensions~\cite{kaw,af,hn,hm1} and SO(10) and E$_6$ models in 6 
dimensions~\cite{abc,hnos,hks,hjll} have recently been constructed. At first 
sight, possible couplings in theories of this type are severely restricted. 
On the one hand, the large gauge and supersymmetry of the bulk excludes 
many couplings. On the other hand, half of the bulk fields are odd under the 
discrete symmetry defining the orbifold and therefore vanish at the boundary
(or brane), where more couplings are allowed. Furthermore, certain degrees 
of freedom are defined as brane fields and can therefore not participate in 
bulk interactions. Superfield brane operators lift many of these restrictions
since, due to the presence of the $x^5$ derivative operator, bulk fields 
that are odd under the discrete symmetry (i.e., vanish at the boundary but 
have non-zero derivative) can participate in brane-localized interactions. 
We briefly survey the relevance of these operators to 5d SUSY GUTs, 
emphasizing, in particular, proton decay operators and brane-localized 
mass terms. 

The paper is organized as follows. After defining the 5d SYM theory in 
Sect.~2, we explicitly derive its 4d superfield formulation in Sect.~3. Here, 
our main result is the simple and manifestly gauge invariant formulation 
based on the covariant derivative in the $x^5$ direction 
(Eqs.~(\ref{nab5})--(\ref{sflag})). Section~4 describes the bulk 
hypermultiplet while Sect.~5 outlines the classification of brane 
operators using the now available gauge covariant formalism. Applications 
to orbifold GUTs are discussed in Sect.~6, followed by the conclusions in 
Sect.~7.

\section{Gauge multiplet in 5 dimensions}
We begin by describing the $N=1$ (8 supercharges) 5d gauge 
multiplet~\cite{mp} (cf. also~\cite{5dss}) using conventions which are as 
close as possible to~\cite{wb}. This will make the following transition to 
4d superfields particularly simple. 

Capitalized indices $M,N,..$ run over $0,1,2,3,5$; lower case indices 
$m,n,..$ run over $0,1,2,3$. The metric is $\eta_{M,N}=$diag$(-1,1,1,1,1)$ 
and the Dirac matrices can be chosen as
\be
\gamma^M=\left(\,\left(\begin{array}{cc}0&\sigma^m\\ \bar{\sigma}^m&0
\end{array}\right),\left(\begin{array}{cc}-i&0\\ 0&i\end{array}\right)\,
\right)\,,
\ee
where $\sigma^m=(1,\vec{\sigma})$ and $\bar{\sigma}^m=(1,-\vec{\sigma})$. 
It is convenient to use symplectic Majorana spinors $\psi^i$, where $i=1,2$ 
transforms under an SU(2) R-symmetry. The reality condition reads
\be
\psi^i=\epsilon^{ij}C\bar{\psi}_j^T\,,
\ee
where the 5d charge conjugation matrix $C$ satisfies $C\gamma^MC^{-1}=
(\gamma^M)^T$. We use the explicit form $C=$diag$(i\sigma^2,i\sigma^2)$. 
Note that lower indices $i,j,..$ transform under the $\bar{2}$ of SU(2)-R 
and the $\epsilon$ tensor ($\epsilon^{12}=\epsilon_{21}=1$) can be used to 
raise or lower indices.

The 5d gauge multiplet contains a vector $A^M$, a real scalar $\Sigma$, and 
an SU(2)-R doublet of gauginos $\lambda^i$. Furthermore, one requires three 
real auxiliary fields $X^a$, which form a triplet of SU(2)-R. These fields 
are all in the adjoint representation of the gauge group. The SUSY parameter 
is a symplectic Majorana spinor $\xi^i$, and the transformation laws are 
given 
by\footnote{
The sign of the last term in Eq.~(\ref{delx}) differs from~\cite{mp} due to 
our opposite sign choice of $A_M$ in the covariant derivative $D_M$. The 
signs of the last two terms in Eq.~(\ref{dell}) appear to genuinely 
disagree with~\cite{mp}. The consistency of the present equations is most 
easily confirmed by checking the 4d part of this 5d SUSY, which is 
worked out explicitly below.}
\bea
\delta_\xi A^M &=& i\bar{\xi}_i\gamma^M\lambda^i\label{dela}\\
\delta_\xi \Sigma &=& i\bar{\xi}_i\lambda^i\\
\delta_\xi \lambda^i &=& \left(\gamma^{MN}F_{MN}+\gamma^M D_M\Sigma
\right)\xi^i+i\left(X^a\sigma^a\right)^i_j\xi^j\label{dell}\\
\delta_\xi X^a &=& \bar{\xi}_i(\sigma^a)^i_j\gamma^M D_M\lambda^j+i
\left[\Sigma,\bar{\xi}_i(\sigma^a)^i_j\lambda^j\right]\,,\label{delx}
\eea
where $\gamma^{MN}=\frac{1}{4}[\gamma^M,\gamma^N]$ and $D_M=\partial_M+iA_M$, 
with appropriate adjoint action of $A_M$ implied. The 5d lagrangian, 
invariant under this SUSY, reads
\be
{\cal L}=\frac{1}{g^2}\left(-\frac{1}{2}\mbox{tr}(F_{MN})^2-\mbox{tr}
(D_M\Sigma)^2-\mbox{tr}\left(\bar{\lambda}_ii\gamma^MD_M\lambda^i\right)+
\mbox{tr}(X^a)^2+\mbox{tr}(\bar{\lambda}_i[\Sigma,\lambda^i])\right)\,.
\label{lag}
\ee

\section{Formulation in terms of 4d superfields}
Orbifold compactifications of the fifth dimension break at least half of 
the $N=1$ 5d SUSY (which corresponds to $N=2$ SUSY from the 4d perspective). 
This is obvious since the full set of 5d SUSY transformations generates 
translations in the $x^5$ direction, and the latter are not a symmetry 
of the orbifold. To make the surviving $N=1$ 4d SUSY manifest, it is 
convenient to consider the decomposition of a 5d symplectic Majorana 
spinor $\psi^i$ into its components (two 4d Weyl spinors $\psi_L$ and 
$\psi_R$) under the 4d Lorentz group. It reads
\be
\psi^1=\left(\begin{array}{cc}(\psi_L)_\alpha\\ (\bar{\psi}_R)^{\dot{\alpha}}
\end{array}\right)\,,\quad
\psi^2=\left(\begin{array}{cc}(\psi_R)_\alpha\\ -(\bar{\psi}_L)^{\dot{\alpha}}
\end{array}\right)\,,\quad
\bar{\psi}_1=\left(\begin{array}{cc}(\psi_R)^\alpha\\ 
(\bar{\psi}_L)_{\dot{\alpha}}\end{array}\right)^T\,,\quad
\bar{\psi}_2=\left(\begin{array}{cc}-(\psi_L)^\alpha\\ 
(\bar{\psi}_R)_{\dot{\alpha}}\end{array}\right)^T\,.
\ee
One can now easily work out the 4d Weyl spinor formulation of 
Eqs.~(\ref{dela})--(\ref{lag}). We assume that the surviving 4d SUSY is 
generated by a set of parameters $\xi^i$ defined by the Weyl spinor 
$\xi_L$, with $\xi_R=0$. For convenience, we explicitly give the 
transformation rules of the component fields under this smaller SUSY,
using 4d Weyl spinors:
\bea
\delta_{\xi_L}A^m &=& i\bar{\xi}_L\bar{\sigma}^m\lambda_L+i\xi_L\sigma^m
\bar{\lambda}_L\label{delam}\\
\delta_{\xi_L}A^5 &=& -\bar{\xi}_L\bar{\lambda}_R-\xi_L\lambda_R\\
\delta_{\xi_L}\Sigma &=& i\bar{\xi}_L\bar{\lambda}_R-i\xi_L\lambda_R\\
\delta_{\xi_L}\lambda_L &=& \sigma^{mn}F_{mn}\xi_L-iD_5\Sigma\xi_L
+iX^3\xi_L\\
\delta_{\xi_L}\lambda_R &=& i\sigma^mF_{5m}\bar{\xi}_L-\sigma^mD_m
\Sigma\bar{\xi}_L+i(X^1+iX^2)\xi_L\\
\delta_{\xi_L}(X^1+iX^2) &=& 2\bar{\xi}_L\bar{\sigma}^mD_m\lambda_R-2i
\bar{\xi}_LD_5\bar{\lambda}_L+i[\Sigma,2\bar{\xi}_L\bar{\lambda}_L]\\
\delta_{\xi_L} X^3 &=& \bar{\xi}_L\bar{\sigma}^mD_m\lambda_L+i
\bar{\xi}_LD_5\bar{\lambda}_R-\xi_L\sigma^mD_m\bar{\lambda}_L
-i\xi_LD_5\lambda_R\nonumber\\
&&+i[\Sigma,(\bar{\xi}_L\bar{\lambda}_R+\xi_L\lambda_R)]\,,\label{delx3}
\eea
where $\sigma^{mn}=\frac{1}{4}(\sigma^m\bar{\sigma}^n-\sigma^n\bar{\sigma}^m
)$. 

Now observe~\cite{mp,agw} that the fields $A_m$, $\lambda_L$ and 
$(X^3-D_5\Sigma)$ transform precisely as the components of a vector 
superfield in Wess-Zumino (WZ) gauge:
\be
V=-\theta\sigma^m\bar{\theta}A_m+i\theta^2\bar{\theta}\bar{\lambda}_L-
i\bar{\theta}^2\theta\lambda_L+\frac{1}{2}\theta^2\bar{\theta}^2\left(X^3-
D_5\Sigma\right)\,.\label{vsf}
\ee
Here we use the conventions of~\cite{wb} for the action of the SUSY 
transformation $\delta_{\xi_L}=\xi_L Q+\bar{\xi}_L\bar{Q}$ on $\theta$ 
and $\bar{\theta}$. Furthermore, in slight deviation from the conventions 
of~\cite{wb}, we define super-gauge transformations of vector ($V$) and 
fundamental-representation chiral ($\Psi$) superfields by
\be
e^{2V}\,\,\to\,\, e^{\Lambda^\dagger}e^{2V}e^\Lambda\qquad\mbox{and}
\qquad\Psi\,\,\to\,\,e^{-\Lambda}\Psi\,,
\ee
where $\Lambda$ is a chiral superfield depending, in general, on $x^5$. 
For what follows, it is important to recall that the transformation 
rules for the WZ-gauge component fields are obtained by the application of 
$\delta_{\xi_L}$ to $V$, followed by a super gauge transformation that
takes $V$ back to WZ gauge. This gauge transformation is specified by 
\be
\Lambda=\sqrt{2}\theta\left(\sqrt{2}\sigma^m\bar{\xi}_LA_m\right)+
\theta^2\left(-2i\bar{\xi}_L\bar{\lambda}_L\right)\label{lam}
\ee
in the $y$ basis (i.e., with the component fields as functions of $x^5$ 
and $y^m=x^m+i\theta\sigma^m\bar{\theta}$). 

The next essential observation is that the fields $(\Sigma+iA_5)$, $(-i
\sqrt{2}\lambda_R)$ and $(X^1+iX^2)$ transform as the components of a 
chiral adjoint superfield in the $y$ basis,
\be
\Phi=(\Sigma+iA_5)+\sqrt{2}\theta\left(-i\sqrt{2}\lambda_R\right)+\theta^2
(X^1+iX^2)\,,
\ee
if, at the same time, this field is defined to transform as
\be \Phi\to e^{-\Lambda}(\partial_5+\Phi)e^\Lambda\label{fgau}
\ee
under super gauge transformations. More precisely, this means that the 
transformation rules of Eqs.~(\ref{delam})--(\ref{delx3}) are reproduced 
by first applying $\delta_{\xi_L}$ to $\Phi$ and then returning to WZ gauge 
by a gauge transformation, Eq.~(\ref{fgau}), with $\Lambda$ specified in 
Eq.~(\ref{lam}). This point is essential in deriving the 4d SUSY 
transformation rules directly from the 5d SUSY. 

Given the transformation rule, Eq.~(\ref{fgau}), for $\Phi$, it is clear 
that
\be
\nabla_5\equiv\partial_5+\Phi\label{nab5}
\ee
represents a super gauge covariant derivative in the $x^5$ direction. Thus, 
given a superfield ${\cal O}$, which is in some representation of the 
gauge group and transforms covariantly under super gauge transformations, 
the superfield $\nabla_5{\cal O}$ transforms covariantly in the same way. 
For this it is essential that the Lie-algebra valued field $\Phi$ contained 
in $\nabla_5$ acts on ${\cal O}$ as specified by the representation under
which ${\cal O}$ transforms. In particular, for the real superfield
\be
\nabla_5e^{2V}=\partial_5e^{2V}-\Phi^\dagger e^{2V}-e^{2V}\Phi
\ee
the transformation rule is
\be
\nabla_5e^{2V}\,\,\to\,\,e^{\Lambda^\dagger}\left(\nabla_5e^{2V}\right)
e^\Lambda\,.
\ee

The 5d lagrangian can now be written in 4d superfield language by
combining the two lowest-dimension invariant operators that can be built 
from the superfield $V$ and the covariant derivative\footnote{
Introducing this covariant derivative is crucial for the manifestly gauge 
invariant formulation of the non-abelian lagrangian, which represents a
significant simplification as compared to~\cite{agw,mpo}.} operator 
$\nabla_5$:
\be
{\cal L}=\frac{1}{2g^2}\mbox{tr}\left\{W^\alpha W_\alpha\Big|_{\theta^2}+
\bar{W}_{\dot{\alpha}} \bar{W}^{\dot{\alpha}}\Big|_{\bar{\theta}^2}
+\left(e^{-2V}\nabla_5 e^{2V}\right)^2\Big|_{\theta^2\bar{\theta}^2}
\right\}\,.\label{sflag}
\ee
Here $W_\alpha$ is the field-strength superfield constructed from $V$ in the 
usual way. The above lagrangian reproduces Eq.~(\ref{lag}) up to derivative 
terms. Note that, to achieve this agreement, it is not necessary to integrate 
out the auxiliary fields. 

It will prove convenient to define, by analogy to $W_\alpha$, the 
Lie-Algebra valued superfield $Z=e^{-2V}\nabla_5e^{2V}$. Now the 
lagrangian takes the particularly compact form 
\be
{\cal L}=\frac{1}{2g^2}\mbox{tr}\left\{\,\left(W^\alpha W_\alpha\Big|_{
\theta^2}+\mbox{h.c.}\right)\,+Z^2\Big|_{\theta^2\bar{\theta}^2}\right\}\,.
\label{sflag1}
\ee
The superfield $Z$ is not hermitian, but it satisfies the simple condition 
$Z^\dagger=e^{2V}Ze^{-2V}$. 

Note that it is also possible to turn the argument around and to consider 
the {\it construction} of the 5d SUSY lagrangian on the basis of the 4d 
theory. To achieve this, start with a 4d real superfield $V$ with the 
usual gauge transformation properties. The gauge parameter is the 4d chiral 
superfield $\Lambda$. Now consider both superfields as functions of the 
additional parameter $x^5$. To be able to take derivatives in the $x^5$ 
direction, we are forced to introduce the additional gauge connection $\Phi$, 
which is a 4d chiral superfield depending on $x^5$. The requirement that
$\nabla_5=\partial_5+\Phi$ be a covariant derivative enforces the gauge 
transformation property of Eq.~(\ref{fgau}). It turns out that the two 
lowest-dimension invariant operators that can be built from $V$ and 
$\nabla_5$ add up to the 5d SUSY lagrangian, Eq.~(\ref{sflag}), where the 
relative normalization of the $W^2$ and the $\nabla_5^2$ terms is fixed by 
the requirement of 5d Lorentz covariance. As expected on the basis of the 
4d $N=1$ SUSY and the full 5d Lorentz covariance, the larger 5d $N=1$ SUSY 
emerges as an additional feature. 

Pursuing this line of thinking, it is now straightforward to construct 
higher-derivative operators of the 5d SYM theory in a systematic way. One 
simply has to write down all 4d-SUSY-invariant superfield expressions of 
a given (higher) dimension and constrain the coefficients by the 
requirement of full 5d Lorentz symmetry. 

Though motivated by the idea of orbifold compactification, the discussion 
of this section was so far restricted to the 5d Lorentz invariant theory. 
Once 5d Lorentz invariance is broken and a brane is introduced, the above 
arguments concerning the relative normalization of the $W^2$ and 
$\nabla_5^2$ operators in Eq.~(\ref{sflag}) cease to apply. We adopt the 
attitude that, in the orbifold theory, the bulk lagrangian is nevertheless 
restricted by 5d Lorentz invariance, while brane localized versions of the 
$W^2$ and $\nabla_5^2$ operators with unconstrained relative normalization 
become admissible (see~Sect.~5 for more details). Strictly speaking, one 
would have to appeal to supergravity to put the notion of 5d bulk Lorentz 
symmetry in the presence of a brane on a firm basis.

\section{The hypermultiplet}\label{hyp}
For completeness, we also present the relevant formulae for the 5d matter
multiplet (the hypermultiplet). It contains an SU(2)-R doublet of scalar
fields $H^i$, a Dirac field $\psi$ and a doublet of auxiliary fields $F_i$. 
The transformation laws are (for the ungauged case see~\cite{mp})
\bea
\delta_\xi H^i &=& -\sqrt{2}\epsilon^{ij}\bar{\xi}_j\psi\label{delh}\\
\delta_\xi \psi &=& i\sqrt{2}\gamma^MD_MH^i\epsilon_{ij}\xi^j-\sqrt{2}\Sigma
H^i\epsilon_{ij}\xi^j+\sqrt{2}F_i\xi^i\\
\delta_\xi F_i &=& i\sqrt{2}\bar{\xi}_i\gamma^MD_M\psi+\sqrt{2}\bar{\xi}_i
\Sigma\psi-2i\bar{\xi}_i\lambda^j\epsilon_{jk}H^k\,.\label{delf}
\eea
The off-shell 5d lagrangian, invariant under this SUSY, reads
\bea
{\cal L} &=& -(D_MH)^\dagger_i(D^MH^i)-i\bar{\psi}\gamma^MD_M\psi+
F^{\dagger i}F_i-\bar{\psi}\Sigma\psi+H^\dagger_i(\sigma^aX^a)^i_jH^j
\nonumber\\
&& +H^\dagger_i\Sigma^2H^i+\left(i\sqrt{2}\bar{\psi}\lambda^i\epsilon_{ij}
H^j+\mbox{h.c.}\right)\,.\label{hla}
\eea
In the 4d superfield formulation, the component fields are arranged in the 
two chiral 4d superfields (in the $y$ basis):
\bea
H &=& H^1+\sqrt{2}\theta\psi_L+\theta^2(F_1+D_5H^2-\Sigma H^2)\\
H^c &=& H^\dagger_2+\sqrt{2}\theta\psi_R+\theta^2(-F^{\dagger 2}-D_5
H^\dagger_1-H^\dagger_1\Sigma)\,.
\eea
As in the pure gauge case, the $\xi_L$ part of the transformation laws 
given in Eqs.~(\ref{delh})--(\ref{delf}) follows in the superfield 
formulation by acting with $\delta_{\xi_L}$ on $H$ and $H^c$ and then 
gauge transforming back to WZ gauge. The two superfields gauge transform 
according to $H\to e^{-\Lambda}H$ and $H^c\to H^ce^\Lambda$. The 4d 
superfield expression for the lagrangian, Eq.~(\ref{hla}), reads 
\be
{\cal L}= \bigg(H^\dagger e^{2V}H+H^c e^{-2V}H^{c\dagger}\bigg)\bigg|_{
\theta^2\bar{\theta}^2}+\bigg(H^c\nabla_5 H\Big|_{\theta^2}+\mbox{h.c.}
\bigg)\,.\label{shla}
\ee

\section{Superfield brane operators}
An obvious application of the above 4d superfield formalism is the 
classification of brane operators of a 5d SYM theory compactified to
4d on an orbifold. The most general such orbifold is $I\!\!R^4\times I$, 
where $I$ is an interval parameterized by $x^5=y$ and limited by two 
orbifold fixed points. Without loss of generality, we can discuss a
fixed point at $y=0$ which is left invariant by a $Z_2$ symmetry of the 
5d theory corresponding to the reflection $y\to -y$ of the original 5d 
manifold. 

Consider a $Z_2$ action on the 5d gauge multiplet given by 
\be
V(y)\to PV(-y)P^{-1}\qquad\mbox{and}\qquad \Phi(y)\to -P\Phi(-y)P^{-1}\,.
\label{z2a}
\ee
Here, in the simplest case, $P$ is an element of the gauge group, $P\in G$
(the $Z_2$ acts by inner automorphism). More generally, the transformation 
$V\to PVP^{-1}$ can be replaced by any other automorphism of ${\cal G}
=$Lie($G$) (outer automorphism), under the restriction that the square of 
this automorphism is the identity.

Equation~(\ref{z2a}) is a symmetry of the lagrangian, Eq.~(\ref{sflag}), 
and the sign change of the superfield $\Phi$ is required since $\Phi$ 
enters the lagrangian in combination with $\partial_5$. The fields appearing 
in Eq.~(\ref{sflag1}) transform under the $Z_2$ as $W_\alpha(y)\to P
W_\alpha(-y)P^{-1}$ and $Z(y)\to -PZ(-y)P^{-1}$. 

Given the $Z_2$ action on $V\in{\cal G}$, the Lie algebra ${\cal G}$ can be 
decomposed into its even and odd components, ${\cal G}={\cal H}\oplus{\cal 
H}'$, where ${\cal H}$ generates the subgroup $H\subset G$ preserved by the 
orbifolding. Let $T^a$ and $T^{\hat{a}}$ form a basis of ${\cal H}$ and 
${\cal H}'$ respectively. Then the fields $W_\alpha^a$ and $Z^{\hat{a}}$ are 
even under the $Z_2$ and can have non-zero values at the fixed point, while 
the fields $W_\alpha^{\hat{a}}$ and $Z^a$ are odd and vanish at the fixed 
point. Furthermore, the gauge connection at the boundary is specified by 
$\exp(2V)$, which corresponds to a restriction of the gauge symmetry to $H$
since only $V^a$ is non-vanishing. Thus, the lowest-dimension superfields 
that can appear in brane operators are 
\be
W_\alpha^a\,,\quad (\nabla_5W_\alpha)^{\hat{a}}\,,\quad Z^{\hat{a}}\,,\quad
(\nabla_5Z)^a\,,\label{bsf}
\ee
where the argument $y=0$ is suppressed. Brane operators can be 
constructed from these fields as in a 4d SUSY theory, given the 
restrictions of Lorentz invariance ($W$ is a spinor) and of the 
representation content under $H$. 

We can write down the following quadratic operators:
\bea
{\cal O}_1 &=& c^1_{ab}(W^\alpha)^a(W_\alpha)^b\Big|_{\theta^2}+\mbox{h.c.}
\label{o1}\\
{\cal O}_2 &=& c^2_{a\hat{b}}(W^\alpha)^a(\nabla_5W_\alpha)^{\hat{b}}
\Big|_{\theta^2}+\mbox{h.c.}
\label{o2}\\
{\cal O}_3 &=& c^3_{\hat{a}\hat{b}}(\nabla_5W^\alpha)^{\hat{a}}(\nabla_5
W_\alpha)^{\hat{b}}\Big|_{\theta^2}+\mbox{h.c.}
\label{o3}\\
{\cal O}_4 &=& c^4_{\hat{a}\hat{b}}Z^{\hat{a}}Z^{\hat{b}}\Big|_{\theta^2
\bar{\theta}^2}\label{o4}\\
{\cal O}_5 &=& c^5_{\hat{a}b}Z^{\hat{a}}(\nabla_5Z)^b\Big|_{\theta^2
\bar{\theta}^2}\label{o5}\\
{\cal O}_6 &=& c^6_{ab}(\nabla_5Z)^a(\nabla_5Z)^b\Big|_{\theta^2
\bar{\theta}^2}\,,\label{o6}
\eea
where the $c^i$ are invariant under $H$. The operators ${\cal O}_1$ and 
${\cal O}_4$ have structures that are already present in the 5d lagrangian. 
Nevertheless, as will be discussed in more detail in the next section, even 
they give rise to distinctive new effects if they are included in the 
action in this brane-localized version. The operators ${\cal O}_2$ and 
${\cal O}_5$ depend on the non-trivial condition that invariants $c^2$ and 
$c^5$ with mixed indices exist. To see that this is possible in principle, 
consider a group $G=$ U(1)$\times$U(1), broken by outer automorphism of one 
of the U(1)s to $H=$ U(1). In this case, both $W$ and $Z$ are singlets and 
all of the above operators can be present. 

Note also that, if a chiral superfield $\Psi$ is localized at the fixed 
point, this field can be gauged under the group $H$. In this case, the 
kinetic term of this field has the form $\Psi^\dagger\exp(2V^aT^a)\Psi$, 
with $T^a$ in the representation appropriate to $\Psi$. 

Next, consider the case where hypermultiplets (cf.~Sect.~\ref{hyp}) are also
present in the bulk. For a hypermultiplet $(H,H^c)$, a $Z_2$ action 
consistent with Eq.~(\ref{z2a}) is given by
\be
H(y)\to PH(-y)\qquad\mbox{and}\qquad H^c(y)\to -H^c(-y)P\,,\label{htra}
\ee
where the prefactor $-1$ could also be assigned to $H$ instead of $H^c$. 
Choosing a basis in representation space where $H^r$ is even and 
$H^{\hat{r}}$ is odd, the lowest-dimension superfields that can be used for 
brane operators at $y=0$ are
\be
H^r\,,\quad (\nabla_5 H)^{\hat{r}}\,,\quad H^{c\,\hat{r}}\,,\quad (\nabla_5 
H^c)^r\,.
\ee
They can now be combined among each other, with the fields of Eq.~(\ref{bsf}), 
and with chiral brane fields to form invariant brane operators. 
Furthermore, chiral brane fields can be coupled to the fields of 
Eq.~(\ref{bsf}). We do not attempt a complete listing of even the 
lowest-dimensional operators but content ourselves with three examples
that will be useful in the next section: 
\bea
{\cal O}_7 &=& H^c\nabla_5 H\Big|_{\theta^2}+\mbox{h.c.}\label{o7}\\
{\cal O}_8 &=& (\nabla_5H^c) H\Big|_{\theta^2}+\mbox{h.c.}\label{o8}\\
{\cal O}_9 &=& \Psi_1^\dagger e^{2V}(Z^{\hat{a}}T^{\hat{a}})\Psi_2
\Big|_{\theta^2\bar{\theta}^2}+\mbox{h.c.}\,.\label{o9}
\eea
The operators ${\cal O}_7$ and ${\cal O}_8$ are brane-localized versions 
of an operator appearing in the bulk hypermultiplet lagrangian, while 
${\cal O}_9$ couples the gauge fields in the broken directions to chiral 
brane fields $\Psi_1$ and $\Psi_2$ in appropriate representations of the 
subgroup $H$. 

The existence of brane operators of the type discussed in this section 
can have significant impact on the low-energy theory emerging below the 
compactification scale.

\section{Applications}
Recall first the generic setup of 5d orbifold GUTs~\cite{kaw,af,hn,kk,hm1,
hm2}\footnote{
Of course, following the early work on symmetry breaking by 
compactification~\cite{ss,hos}, orbifolds~\cite{dhvw}, and their 
interrelation~\cite{int}, this type of model building was extensively
studied in the framework of string theory. Nevertheless, the present, 
purely field-theoretic constructions are well motivated as attempts to 
compare the wealth of low-energy data with the many possible GUT structures 
in a way that is as direct and simple as possible. For a more detailed 
discussion of the structure of GUT breaking by field-theoretic orbifolding 
see, e.g.,~\cite{hm2}.}. 
One starts with a gauge theory on $I\!\!R^4\times S^1$ and restricts the 
field space by the requirement of symmetry under the discrete group 
$Z_2\times Z_2'$. The $S^1$ is parameterized by $x^5=y\in [0,2\pi R)$ 
or, equivalently, by $y'=y-\pi R/2$. The $Z_2$ action is given by
Eq.~(\ref{z2a}), while the $Z_2'$ action is given by an analogous equation 
where $y$ is replaced by $y'$ and $P$ is replaced by $P'$. Choosing the 
gauge group SU(5) and representation matrices $P=1_5$ and $P'=$ diag$(1,1, 
1,-1,-1)$, one finds that the surviving symmetries on the $P$ and the $P'$ 
branes are SU(5) and $G_{SM}=$SU(3)$\times$SU(2)$\times$U(1). The low-energy 
spectrum of the gauge sector is precisely that of the MSSM since the $P$ 
reflection removes all zero modes from $\Phi$, and the additional $P'$ 
reflection removes the zero modes corresponding to $X,Y$ gauge bosons from 
$V$. To solve the doublet-triplet splitting problem, the Higgs multiplets 
have to be localized in the bulk~\cite{kaw} or on the SM brane~\cite{hm1}. 
Fermions can be placed on the SU(5) brane~\cite{kaw,af,hn}, on the SM 
brane~\cite{hm1}, or in the bulk~\cite{hn,hm1}. 

We now want to briefly discuss several possible implications of brane 
operators localized at the two fixed points for the low-energy theory 
derived from the 5d orbifold GUT. 

First, consider proton decay mediated by the $X,Y$ gauge bosons, which have 
GUT scale masses because their Kaluza-Klein (KK) spectrum does not contain a 
zero mode. Naively one would think that this type of process is absent in 
models where the fermions are localized on the SM brane~\cite{hm1} because 
the $V$ components corresponding to the broken direction vanish at this brane.
However, operators of the type ${\cal O}_9$ in Eq.~(\ref{o9}) may be present,
in which case all the usual couplings of SM particles to the 5d analogues of
$X,Y$ gauge bosons may exist. Of course, the coupling strength is now not 
any more an unambiguous prediction of the theory. We leave the more detailed 
investigation of proton decay in this and other scenarios to a future
publication~\cite{hm3}. 

Next, consider the masses of the Higgs fields. It is one of the most 
attractive features of the present models that, after appropriate orbifold 
projections, a bulk hypermultiplet $(H,H^c)$ in the 5 of SU(5) gives rise to 
one doublet chiral superfield. More specifically, this is realized by using 
Eq.~(\ref{htra}) as it stands for the $Z_2$ transformation and switching 
the prefactor $-1$ from $H^c$ to $H$ for the $Z_2'$ transformation. However, 
given the presence of the operator ${\cal O}_8$ in Eq.~(\ref{o8}) on one of 
the branes, one obtains a mixing between the doublet zero mode from $H$ and 
the massive KK modes of the doublet from $H^c$. 

To see this in more detail, consider explicitly the part of the full action 
that is quadratic in fermionic fields and does not include derivatives in 
brane-parallel directions. We integrate the lagrangian (Eq.~(\ref{hla}) or 
Eq.~(\ref{shla})) over $y\in[0,\pi R/2]$ (where $R\sim 1/M_{GUT}$ is the 
compactification scale), add $c\,{\cal O}_8$ at the $P$ brane, and restrict 
our attention to the SU(2) doublet components. The relevant terms read:
\be
S=\int d^4x\int_0^{\pi R/2}dy\left(\,[1-c\delta(y)]\psi_L\partial_5\psi_R
+\mbox{h.c.}\,\right)+\cdots\,.
\ee
This action has to be varied under the constraints that $\psi_R$ and 
$\partial_5\psi_L$ vanish at the boundaries. The resulting equations of 
motion are
\bea
\partial_5\left[(1-c\delta(y))\psi_L\right] &=& 0\\
(1-c\delta(y))\partial_5\psi_R &=& 0\,.
\eea
Thus, for $c\neq 0$, the usual zero mode, $\psi_L=$ const., $\psi_R=0$, is 
removed. Formally, one finds a modified zero mode, $\psi_L\sim[1-c 
\delta(y)]^{-1}$, $\psi_R=0$. However, this is a highly singular function 
which may couple strongly via various higher derivative operators. 
Therefore, even though we can not exclude the existence of a related 
zero-mode in the UV completion of the theory, it does not seem to be an 
unambiguous prediction of the low-energy effective theory. In fact, when 
integrating out the auxiliary fields to analyse the scalar part of the 
action, one finds singular contributions reminiscent of the infamous 
$\delta(0)$ terms discussed in~\cite{mp}. Note also that solutions arising 
in the presence of different types of brane-localized operators 
have been discussed, e.g., in~\cite{bl}. 

Thus, given the limitations of our leading-order, purely field-theoretic 
analysis, we conclude that the operator ${\cal O}_8$ significantly affects 
the Higgs zero mode. Even if a modified zero mode should still be present, 
its strong suppression at the brane may cause problems for the (necessarily 
brane-localized) Yukawa interactions. Fortunately, this operator is protected
from quantum corrections and may therefore safely be set to zero at a 
technical level. However, one may still be concerned by the fact that, due 
to the presence of this operator in the bulk action, there is no obvious 
symmetry argument excluding the brane-localized version. 

Now we turn to the pure gauge sector. The operator ${\cal O}_1$ of 
Eq.~(\ref{o1}) has already been extensively discussed in the context of 
orbifold GUTs. In the case where $G=$ SU(5) and $H=$ SU(3)$\times$SU(2)$
\times$U(1), it contains three independent pieces which represent the (so 
far uncalculable) threshold corrections of the model~\cite{hn}. The 
logarithmic running of differences of gauge couplings above the 
compactification scale~\cite{nsw} can be understood as the running of the 
coefficients of these operators~\cite{hm1,cprt} (see~\cite{hnuni} for a 
recent more detailed analysis). 

The operator ${\cal O}_4$ of Eq.~(\ref{o4}) has so far not been used in the 
construction of orbifold GUTs. In fact, without the present, fully 
gauge-covariant superfield formalism it is difficult to even write this 
operator down. We now discuss an interesting and, naively, somewhat 
mysterious feature of this operator. For simplicity, let $c^4_{\hat{a}
\hat{b}}=c\delta_{\hat{a}\hat{b}}$ and focus on the quadratic term that 
mixes the broken-direction modes of $\Phi$ and $V$:
\vspace*{-.1cm}

\be
{\cal O}_4=-4c\delta_{\hat{a}\hat{b}}(\Phi+\Phi^\dagger)^{\hat{a}}\partial_5
V^{\hat{b}}\Big|_{\theta^2\bar{\theta}^2}+\cdots\,.\label{qo}
\ee
\vspace*{-.2cm}

\noindent
Now consider a situation where both $P$ and $P'$ act non-trivially in 
group space and focus on fields $\Phi^{\hat{a}}$ and $V^{\hat{a}}$ which 
correspond to a Lie algebra generator broken on both boundaries. In this 
case $\Phi^{\hat{a}}$ has a zero mode. On the one hand, Eq.~(\ref{qo}) 
appears to imply that this zero mode is lifted by mixing with the massive 
KK modes of $V^{\hat{a}}$. On the other hand, this zero mode corresponds 
to the freedom one has in choosing the relative orientation of the 
symmetry groups on the two branes as subgroups of $G$. This modulus can 
be described by the Wilson line connecting the two boundaries\footnote{
The vanishing tree-level potential for this degree of freedom is protected 
by SUSY but can receive radiative corrections in non-SUSY 
theories~\cite{kly}.}. The 
latter can clearly take a non-trivial value by having a gauge potential 
$A_5$ that vanishes near both branes and is non-zero only in the middle 
of the bulk. Thus, it should be unaffected by brane operators. This 
apparent contradiction is resolved by recalling the inhomogeneous gauge 
transformation property of $\Phi$. In fact, $\Phi$ can always be gauged to 
zero at the brane. An appropriate gauge transformation parameter $\Lambda$ 
is defined by 
\be
\partial_5e^{\Lambda}=-\Phi e^{\Lambda}\,\,\,,\qquad 
\Lambda\big|_{\mbox{\footnotesize brane}}
=0
\ee
(which is clearly consistent with broken gauge invariance at the brane). 
This explains why Eq.~(\ref{qo}) can not be used to argue that 
$\Phi^{\hat{a}}$ obtains a mass\footnote{
A physical situation where these considerations apply arises if one attempts 
to construct a 5d SO(10) model by breaking the group to SU(5)$\times$U(1) 
and SU(5)$'\times$U(1)$'$ on the two boundaries. Although the intersection 
of these groups is $G_{SM}\times$U(1), the model is plagued by the presence 
of $\Phi$ zero modes~\cite{hnos}. The above considerations show that this 
problem is rather fundamental and can not be overcome by brane 
operators.}.

We leave the discussion of other operators and their role in specific 
models to future, more phenomenologically oriented work.

\section{Conclusions}
In this paper, we have given a detailed derivation of the 4d superfield 
formulation of a 5d SYM theory compactified on a field-theoretic orbifold. 
The 4d SUSY has been explicitly identified as the unbroken part of the 
larger SUSY of the original 5d theory. An essential ingredient of our 
treatment is the gauge and supersymmetry covariant derivative in the $x^5$ 
direction, $\nabla_5=\partial_5+\Phi$. The Lie-Algebra valued chiral 
superfield $\Phi$ represents the gauge connection in the $x^5$ direction. Its
action in field space is specified by the usual Lie algebra action on fields 
in a representation of the gauge group. The recognition of the full 
covariance of $\nabla_5$ and the resulting simplification of the 4d 
superfield formulation in the non-abelian case 
(cf.~Eqs.~(\ref{nab5})--(\ref{sflag})) represents our main 
conceptual progress compared to the earlier treatment of~\cite{agw}. 
An immediate consequence is the possibility to construct higher-derivative 
operators in the 5d theory by combining terms with 4d covariant 
derivatives and $\nabla_5$ under the restriction of full 5d Lorentz 
invariance. 

In our formulation, it is straightforward to write down brane operators 
localized at orbifold fixed points where the gauge symmetry is broken to 
a subgroup of the original symmetry group (cf.~Eqs.~(\ref{bsf})--(\ref{o9})).
This has particular relevance for the phenomenology of orbifold GUT models. 
It is now possible to discuss brane-localized couplings of fields that 
vanish at the brane. This is achieved using the $\nabla_5$ derivative of 
those fields, which is in general non-zero at the brane. 

One implication is the possibility of proton decay mediated by $X,Y$ gauge 
bosons even in the case where fermionic matter is localized at a fixed point 
where the gauge symmetry is restricted to the standard model group. Another 
implication is the possibility of a brane-localized mass term mixing the 
light Higgs doublet with the heavy Kaluza-Klein modes from the 5d 
hypermultiplet. A further potential area of application, which has not 
been discussed here but where brane operators may play an important role, 
is the low-energy supersymmetry breaking in models with extra dimensions 
(see, e.g.,~\cite{ssb}).

We hope that the developed framework will prove useful in the detailed 
phenomenological analysis of different specific orbifold GUT models. It 
would furthermore be important to generalize the presented gauge-covariant 
treatment to SYM theories in more than 5 dimensions. 

\vspace*{1cm}

\noindent
{\bf Acknowledgments}:
I am very grateful to J. March-Russell for numerous detailed discussions 
at various stages of this project. I would also like to thank S. Ferrara, 
R. Rattazzi and C. Scrucca for helpful conversations.


\begin{thebibliography}{99}

\bibitem{kaw}  Y.~Kawamura, Prog.\ Theor.\ Phys.\  {\bf 105} (2001) 999 
               [arXiv:hep-ph/0012125].

\bibitem{af}   G.~Altarelli and F.~Feruglio, Phys.\ Lett.\ B {\bf 511} 
               (2001) 257 [arXiv:hep-ph/0102301].

\bibitem{hn}   L.~Hall and Y.~Nomura, Phys.\ Rev.\ D {\bf 64} (2001) 055003 
               [arXiv:hep-ph/0103125].

\bibitem{kk}   T.~Kawamoto and Y.~Kawamura, arXiv:hep-ph/0106163.

\bibitem{hm1}  A.~Hebecker and J.~March-Russell, Nucl.\ Phys.\ B {\bf 613} 
               (2001) 3\\{}[arXiv:hep-ph/0106166].

\bibitem{hm2}  A.~Hebecker and J.~March-Russell, arXiv:hep-ph/0107039.

\bibitem{abc}  T.~Asaka, W.~Buchm\"uller and L.~Covi, arXiv:hep-ph/0108021.

\bibitem{hnos} L.~J.~Hall, Y.~Nomura, T.~Okui and D.~R.~Smith, 
               arXiv:hep-ph/0108071.

\bibitem{cprt} R.~Contino, L.~Pilo, R.~Rattazzi and E.~Trincherini,
               arXiv:hep-ph/0108102.

\bibitem{hjll} C.~S.~Huang, J.~Jiang, T.~Li and W.~Liao, arXiv:hep-th/0112046.

\bibitem{hks}  N. Haba, T. Kondo and Y. Shimizu, arXiv:hep-ph/0112132. 

\bibitem{oth}  A.~B.~Kobakhidze, Phys.\ Lett.\ B {\bf 514} (2001) 131
               [arXiv:hep-ph/0102323];\\
               R.~Barbieri, L.~J.~Hall and Y.~Nomura, arXiv:hep-th/0107004;\\
               J.~A.~Bagger, F.~Feruglio and F.~Zwirner, 
               arXiv:hep-th/0107128;\\
               T.~j.~Li, Phys.\ Lett.\ B {\bf 520} (2001) 377
               [arXiv:hep-th/0107136],\\ arXiv:hep-ph/0108120,
               arXiv:hep-ph/0108238, and arXiv:hep-th/0110065;\\
               N.~Haba, Y.~Shimizu, T.~Suzuki and K.~Ukai, 
               arXiv:hep-ph/0107190;\\
               A.~Masiero, C.~A.~Scrucca, M.~Serone and L.~Silvestrini,
               arXiv:hep-ph/0107201;\\
               C.~A.~Scrucca and M.~Serone, JHEP {\bf 0110} (2001) 017
               [arXiv:hep-th/0107159];\\
               L.~J.~Hall, H.~Murayama and Y.~Nomura, arXiv:hep-th/0107245;\\
               C.~Csaki, G.~D.~Kribs and J.~Terning, arXiv:hep-ph/0107266;\\
               H.~C.~Cheng, K.~T.~Matchev and J.~Wang, Phys.\ Lett.\ B 
               {\bf 521} (2001) 308\\{} [arXiv:hep-ph/0107268];\\
               N.~Maru, Phys.\ Lett.\ B {\bf 522} (2001) 117
               [arXiv:hep-ph/0108002];\\
               N.~Haba, T.~Kondo, Y.~Shimizu, T.~Suzuki and K.~Ukai,
               arXiv:hep-ph/0108003;\\
               L.~Hall, J.~March-Russell, T.~Okui and D.~R.~Smith, 
               arXiv:hep-ph/0108161;\\
               N.~Borghini, Y.~Gouverneur and M.~H.~Tytgat, 
               arXiv:hep-ph/0108094;\\
               M.~Chaichian, J.~L.~Chkareuli and A.~Kobakhidze, 
               arXiv:hep-ph/0108131;\\
               R.~Dermisek and A.~Mafi, arXiv:hep-ph/0108139;\\
               T.~Watari and T.~Yanagida, Phys.\ Lett.\ B {\bf 519} (2001) 
               164 [arXiv:hep-ph/0108152];\\
               Y.~Nomura, arXiv:hep-ph/0108170;\\
               Q.~Shafi and Z.~Tavartkiladze, arXiv:hep-ph/0108247;\\
               N.~Haba, arXiv:hep-ph/0110164;\\
               G.~A.~Diamandis, B.~C.~Georgalas, P.~Kouroumalou and 
               A.~B.~Lahanas,\\ arXiv:hep-th/0111046;\\
               H.~D.~Kim, J.~E.~Kim and H.~M.~Lee, arXiv:hep-ph/0112094.

\bibitem{ant}  I.~Antoniadis, Phys.\ Lett.\ B {\bf 246} (1990) 377;\\
               G.~R.~Dvali and M.~A.~Shifman, Nucl.\ Phys.\ B {\bf 504} 
               (1997) 127\\{} [arXiv:hep-th/9611213];\\
               A.~Pomarol and M.~Quiros, Phys.\ Lett.\ B {\bf 438} (1998) 255
               [arXiv:hep-ph/9806263]. 

\bibitem{ddg}  K.~R.~Dienes, E.~Dudas and T.~Gherghetta, Phys.\ Lett.\ B 
               {\bf 436} (1998) 55\\{}[arXiv:hep-ph/9803466] and Nucl.\ 
               Phys.\ B {\bf 537} (1999) 47 [arXiv:hep-ph/9806292].

\bibitem{mss}  N.~Marcus, A.~Sagnotti and W.~Siegel, Nucl.\ Phys.\ B 
               {\bf 224} (1983) 159.

\bibitem{agw}  N.~Arkani-Hamed, T.~Gregoire and J.~Wacker, 
               arXiv:hep-th/0101233.

\bibitem{mpo}  D.~Marti and A.~Pomarol, Phys.\ Rev.\ D {\bf 64} (2001) 105025
               [arXiv:hep-th/0106256].

\bibitem{mp}   E.~A.~Mirabelli and M.~E.~Peskin, Phys.\ Rev.\ D {\bf 58} 
               (1998) 065002\\{} [arXiv:hep-th/9712214].

\bibitem{5dss} E. Cremmer, {\it Superspace and Supergravity}, proceedings, 
               Eds. S.W. Hawking and M. Rocek., Cambridge Univ. Press, 1981;\\
               M.~Gunaydin, G.~Sierra and P.~K.~Townsend, Nucl.\ Phys.\ B 
               {\bf 242} (1984) 244;\\
               A.~Salam and E.~Sezgin, {\it Supergravities in Diverse 
               Dimensions}, Vols. 1 and 2, World Scientific, 1989;\\
               E.~R.~Sharpe, Nucl.\ Phys.\ B {\bf 523} (1998) 211
               [arXiv:hep-th/9611196].

\bibitem{wb}   J. Wess and J. Bagger, {\it Supersymmetry and Supergravity},
               Princeton Univ. Press, 1983

\bibitem{ss}   J.~Scherk and J.~H.~Schwarz, Phys.\ Lett.\ B {\bf 82} (1979) 
               60 and Nucl.\ Phys.\ B {\bf 153} (1979) 61.

\bibitem{hos}  Y.~Hosotani, Phys.\ Lett.\ B {\bf 126} (1983) 309 and 
               Annals Phys.\ {\bf 190} (1989) 233.

\bibitem{dhvw} L.~Dixon, J.~A.~Harvey, C.~Vafa and E.~Witten, Nucl.\ Phys.\ 
               B {\bf 261} (1985) 678 and Nucl.\ Phys.\ B {\bf 274} (1986) 
               285.

\bibitem{int}  L.~E.~Ibanez, H.~Nilles and F.~Quevedo, Phys.\ Lett.\ B 
               {\bf 187} (1987) 25;\\
               L.~E.~Ibanez, J.~Mas, H.~Nilles and F.~Quevedo, Nucl.\ Phys.\ 
               B {\bf 301} (1988) 157. 

\bibitem{hm3}  A.~Hebecker and J.~March-Russell, work in progress.

\bibitem{bl}   Z.~Chacko, M.~A.~Luty and E.~Ponton, JHEP {\bf 0007} (2000) 
               036\\{} [arXiv:hep-ph/9909248];\\
               N.~Arkani-Hamed et al., Nucl.\ Phys.\ B {\bf 605} (2001) 81
               [arXiv:hep-ph/0102090];\\
               D.~E.~Kaplan and T.~M.~Tait, JHEP {\bf 0111} (2001) 051
               [arXiv:hep-ph/0110126].

\bibitem{nsw}  Y.~Nomura, D.~R.~Smith and N.~Weiner, Nucl.\ Phys.\ B 
               {\bf 613} (2001) 147\\{} [arXiv:hep-ph/0104041].

\bibitem{hnuni}L.~J.~Hall and Y.~Nomura, arXiv:hep-ph/0111068.

\bibitem{kly}  M.~Kubo, C.~S.~Lim and H.~Yamashita, arXiv:hep-ph/0111327.

\bibitem{ssb}  L.~Randall and R.~Sundrum, Nucl.\ Phys.\ B {\bf 557} (1999) 
               79 [arXiv:hep-th/9810155];\\
               D.~E.~Kaplan, G.~D.~Kribs and M.~Schmaltz, Phys.\ Rev.\ D 
               {\bf 62} (2000) 035010\\{} [arXiv:hep-ph/9911293];\\
               Z.~Chacko, M.~A.~Luty, A.~E.~Nelson and E.~Ponton,
               JHEP {\bf 0001} (2000) 003 [arXiv:hep-ph/9911323];\\
               M.~Luty and R.~Sundrum, arXiv:hep-th/0111231;\\
               A.~Anisimov, M.~Dine, M.~Graesser and S.~Thomas,
               arXiv:hep-th/0111235.
\end{thebibliography}
\end{document}